\documentclass[aps,pre,twocolumn,floats,showpacs]{revtex4}

\usepackage{epsfig}
\usepackage{subfigure}
\usepackage{color}
\usepackage{bm}
\usepackage{latexsym}
\usepackage{hyperref}

\begin{document}
\newcommand{\hide}[1]{}
\newcommand{\tbox}[1]{\mbox{\tiny #1}}
\newcommand{\half}{\mbox{\small $\frac{1}{2}$}}
\newcommand{\sinc}{\mbox{sinc}}
\newcommand{\const}{\mbox{const}}
\newcommand{\trc}{\mbox{trace}}
\newcommand{\intt}{\int\!\!\!\!\int }
\newcommand{\ointt}{\int\!\!\!\!\int\!\!\!\!\!\circ\ }
\newcommand{\eexp}{\mbox{e}^}
\newcommand{\bra}{\left\langle}
\newcommand{\ket}{\right\rangle}
\newcommand{\EPS} {\mbox{\LARGE $\epsilon$}}
\newcommand{\ar}{\mathsf r}
\newcommand{\im}{\mbox{Im}}
\newcommand{\re}{\mbox{Re}}
\newcommand{\bmsf}[1]{\bm{\mathsf{#1}}}
\newcommand{\mpg}[2][1.0\hsize]{\begin{minipage}[b]{#1}{#2}\end{minipage}}

\title{Multifractality in random networks with power--law decaying bond strengths}

\author{Didier A. Vega-Oliveros,$^1$ J. A. M\'endez-Berm\'udez,$^2$ and Francisco A. Rodrigues$^3$}
\affiliation{
$^1$Departamento de Computa\c{c}\~{a}o e Matem\'{a}ticas, Faculdade de Filosofia Ci\^{e}ncias e 
Letras de Ribeir\~{a}o Preto, Universidade de S\~{a}o Paulo, CEP 14040-901, Ribeir\~{a}o Preto, SP, Brasil\\
$^2$Instituto de F\'isica, Benem\'erita Universidad Aut\'onoma de Puebla, 
Apartado Postal J-48, 72570 Puebla, M\'exico\\
$^3$Departamento de Matem\'{a}tica Aplicada e Estat\'{i}stica, Instituto de Ci\^{e}ncias Matem\'{a}ticas 
e de Computa\c{c}\~{a}o, Universidade de S\~{a}o Paulo - Campus de S\~{a}o Carlos, CP 668,
13560-970 S\~{a}o Carlos, SP, Brasil
}

\date{\today}

\begin{abstract}
In this paper we demonstrate numerically that random networks whose adjacency matrices ${\bf A}$ 
are represented by a diluted version of the Power--Law Banded Random Matrix (PBRM) model have 
multifractal eigenfunctions. 
The PBRM model describes one--dimensional samples with random long--range bonds.
The bond strengths of the model, which decay as a power--law, 
are tuned by the parameter $\mu$ as $A_{mn}\propto |m-n|^{-\mu}$; while the 
sparsity is driven by the average network connectivity $\alpha$: for $\alpha=0$ 
the vertices in the network are isolated and for $\alpha=1$ the network is fully 
connected and the PBRM model is recovered. Though it is known that the PBRM model has
multifractal eigenfunctions at the critical value $\mu=\mu_c=1$, we clearly 
show [from the scaling of the relative fluctuation of the participation number 
$I_2$ as well as the scaling of the probability distribution functions $P(\ln I_2)$] 
the existence of the critical value $\mu_c\equiv \mu_c(\alpha)$ for $\alpha<1$.
Moreover, we characterise the multifractality of the eigenfunctions of our
random network model by the use of the corresponding multifractal dimensions $D_q$,
that we compute from the finite network-size scaling of the typical eigenfunction 
participation numbers $\exp\left\langle\ln I_q \right\rangle$.
\end{abstract}

\pacs{64.60.Aq		
      89.75.Hc		
}

\maketitle


\section{Introduction}

Fractality is related to phase transitions in critical phenomena observed in several complex systems~\cite{Stanley71, Vicsek92}. Blood vessels, proteins, ocean waves, animal collaboration patterns, and earthquakes exhibit fractality~\cite{Mandelbrot1983, Bunde013}. Fractality can also be understood as a signature of the organisation and structure of complex systems, which is far from random or regular~\cite{Costa011}. Moreover, the structure of complex systems can be mapped to networks, whose structure~\cite{SHM05} and evolution~\cite{Song06} exhibit fractal properties. 

Fractality in networks has been extensively discussed from several perspectives~\cite{SHM05, Costa07}. These studies have focused on the structural characterisation of fractal networks~\cite{SHM05, Song06} or networks expressly constructed as fractal objects (deterministic or disordered, e.g. see~\cite{MB11,GSKK06,CR10,LYZ14,STM17}). In this respect some algorithms have been developed and applied to compute the fractal dimension of complex networks, see for example~\cite{SHM05,GSKK06,KGS07,SGHM07,S07,LKBH11,FY11,LYA15} and references therein. On the other hand, given a fractal network, there is plenty of works devoted to the signatures of the fractality on the network properties. Among them we can mention the underlying tree structure or skeleton \cite{GSKK06,KGS07} as well as dynamical and transport properties, see for example \cite{MB11,Y02,GSH07,ZZTC08,TL99,ZYG11,KDM15}.

Here, we approach an alternative but close-related subject: We explore the fractality of the eigenfunctions of the adjacency matrices ${\bf A}$ of a random network model. Moreover, we demonstrate  that imposing power--law correlations, i.e.~$A_{mn}\propto |m-n|^{-\mu}$ with  $\mu\sim 1$, on a random network model of the Erd\"os-R\'enyi--type produces multifractal eigenfunctions.

Therefore, in the following section we first review the Power--Law Banded Random Matrix (PBRM) model; a random matrix model used to study the Anderson metal-to-insulator phase transition, which presents multifractal eigenfunctions at the transition point. Then, we introduce the diluted PBRM (dPBRM) model as an ensemble of adjacency matrices of random networks of the Erd\"os-R\'enyi--type.
Using scaling arguments, in Sect.~III we show that the dPBRM model also exhibits a metal-to-insulator phase transition where the corresponding eigenfunctions are multifractal objects. Finally, in Sect.~IV we draw our conclusions.

\section{The random network model}

\subsection{The Power--Law Banded Random Matrix model}

The Power--Law Banded Random Matrix (PBRM) model \cite{MFDQS96} is represented by 
$N\times N$ real symmetric matrices whose elements are 
statistically independent random variables drawn from a normal distribution with 
zero mean, $\langle A_{mn} \rangle=0$, and a variance given by 
\begin{equation}
   \langle |A_{mn}|^2 \rangle =
   \frac{1}{2} \left( \frac{1+\delta_{mm}}{1+\left[
   \sin\left( \pi|m-n|/N \right)/(\pi b/N) \right]^{2\mu}} \right) \ ,
\label{PBRMp}
\end{equation}
where $b$ and $\mu$ are the model parameters. The PBRM model has been 
used to describe one--dimensional tight-binding wires of length $N$ with random 
long--range hoppings. In Eq.~(\ref{PBRMp}) the PBRM model is in its periodic version; 
i.e.,~the one--dimensional wire is in a ring geometry. Theoretical considerations 
\cite{MFDQS96,EM08,Mirlin00,KT00} and detailed numerical investigations 
\cite{EM08,EM00b,V00,V02} have verified that the PBRM model undergoes a transition at 
$\mu=1$, i.e., from localized eigenfunctions for $\mu>1$ to delocalized eigenfunctions for $\mu<1$. 
This transition shows key features of the disorder driven Anderson metal-insulator 
transition \cite{EM08,MKV05,APB09,MAV14}, including multifractality of eigenfunctions and non-trivial 
spectral statistics. Thus the PBRM model possesses a line of critical points 
$b\in (0,\infty)$ at $\mu=\mu_c=1$. By tuning the parameter $b$, from $b\ll 1$ to $b\gg 1$, the 
eigenfunctions cross over from the strong multifractality ($D_q\sim b \to 0$) which 
corresponds to localized--like or insulator--like eigenfunctions to weak multifractality 
($D_q \to 1$), showing rather extended, i.e.,~metallic--like eigenfunctions \cite{EM08,Mirlin00}. 
Here, $D_q$ are the eigenfunction's multifractal dimensions (to be defined in Sect.~\ref{sectIII}). 
At the true Anderson transition in $d=3$ or at the integer 
quantum--Hall transition in $d=2$ the eigenfunctions belong to the weakly multifractal regime, 
i.e.,~$d-D_2\ll d$; it is relevant to note that the PBRM model allows for investigations without such a 
limitation.

\subsection{The diluted Power--Law Banded Random Matrix model}

Here we introduce the 
{\it diluted} PBRM (dPBRM) model as follows: Starting with the PBRM model, we randomly set off-diagonal matrix elements to zero such that the sparsity (i.e.,~the average network connectivity) $\alpha$ is defined as the fraction of the $N(N-1)/2$ independent non-vanishing off-diagonal matrix elements. According to this definition, a diagonal random matrix is obtained for $\alpha = 0$, whereas the PBRM model is recovered when $\alpha = 1$. 

The Erd\"os-R\'enyi adjacency matrix is considered as a mask to define the nonzero
matrix elements of our dPBRM model. Hence, notice that the dPBRM model of size $N$ works as an ensemble of adjacency matrices of Erd\"os-R\'enyi--type networks formed by $N$ vertices. For such networks we allow self-edges and further consider all edges to have random strengths; however, notice that the random strengths are power--law modulated, see Eq.~(\ref{PBRMp}). 

The power--law correlations of the dPBRM model are 
tuned by the parameter $\mu$ as $A_{mn}\propto |m-n|^{-\mu}$, see Eq.~(\ref{PBRMp}).
Notice that for $\mu\to\infty$ the vertices in the
network become isolated since $A_{mn}\to 0$; while for $\mu\to 0$ the dPBRM model 
reproduces the Erd\"os-R\'enyi random network model with {\it maximal disorder} 
(see Refs.~\cite{MAM13,MAM15,GAC18}).
However, here we set $\mu\sim 1$ such that we recover the PBRM model at {\it criticality}
(i.e.,~the PBRM model having multifractal eigenfunctions) when $\alpha=1$.
Moreover, without loss of generality, we will set the effective bandwidth $b$ of 
the dPBRM model to unity; that is, we use the bandwidth that produces multifractal
eigenfunctions with intermediate fractality, $D_2\approx 0.5$, in the PBRM model.
Here $D_2$ is the correlation dimension of the eigenfunctions.

Note that another diluted version of the PBRM model was reported in 
Refs.~\cite{LL01,SAML08,AML08} in studies of quantum percolation.

In the following Section we demonstrate that the eigenfunctions of the dPBRM model are multifractal objects. Besides, we share the implementation and analyses of the reported model online~\footnote{Code available at \url{https://github.com/didiervega/RandomMatrix}}, for easier reproducibility.

\begin{figure*}[!htbp]
\centering	
  \subfigure{\includegraphics[width= 0.31\textwidth]{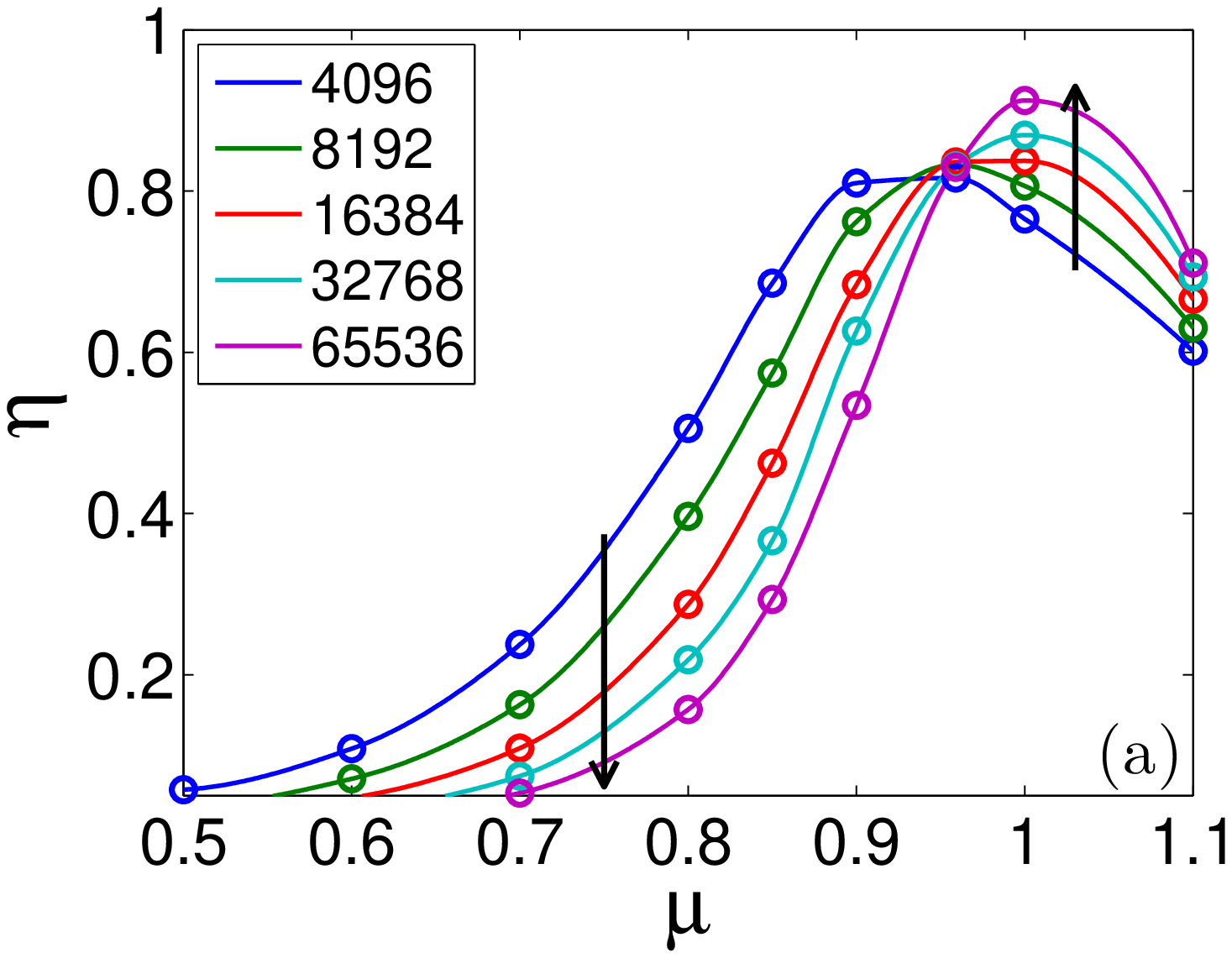}} \hspace{0.3cm}  
  \subfigure{\includegraphics[width= 0.31\textwidth]{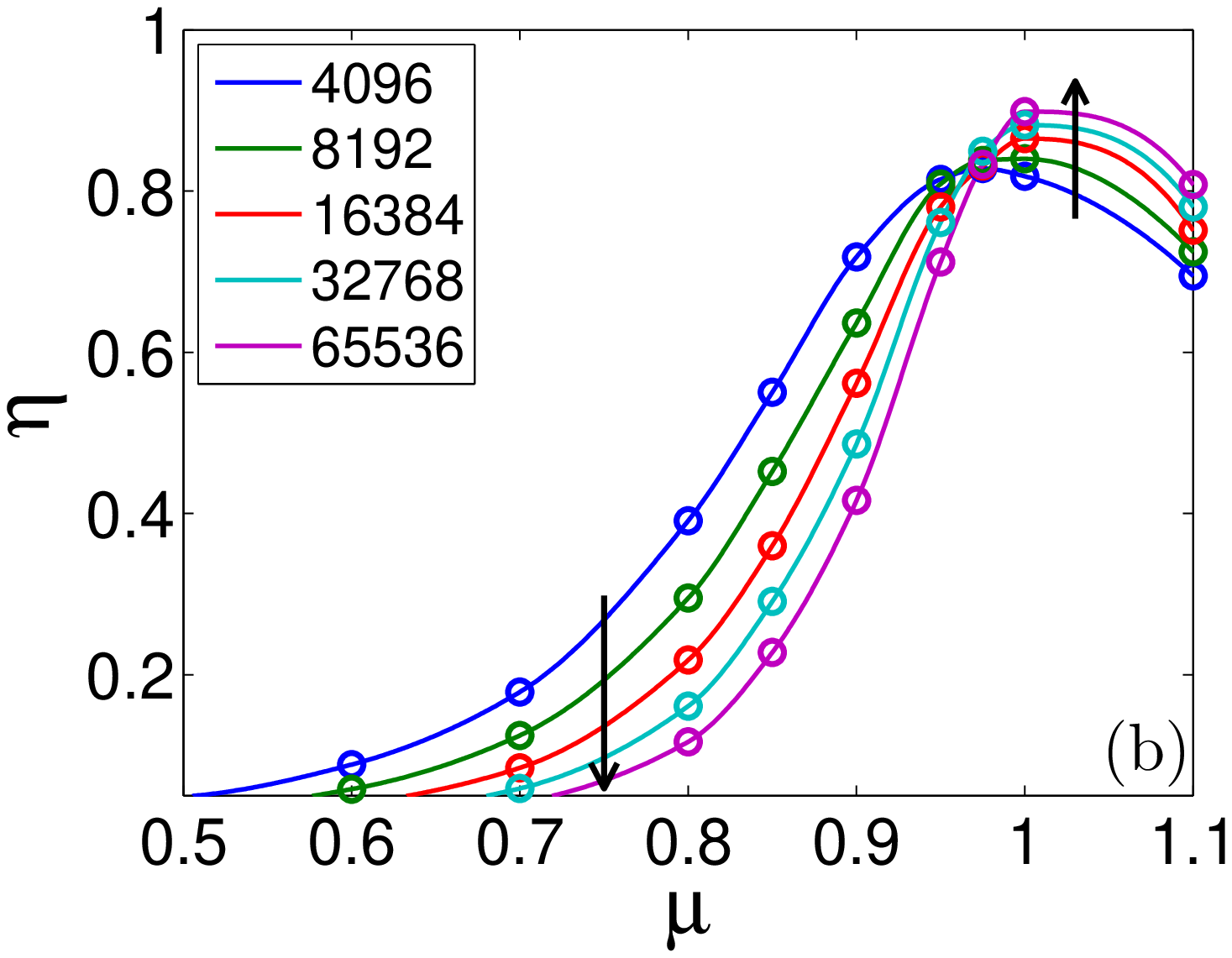}} \hspace{0.3cm} 
  \subfigure{\includegraphics[width= 0.31\textwidth]{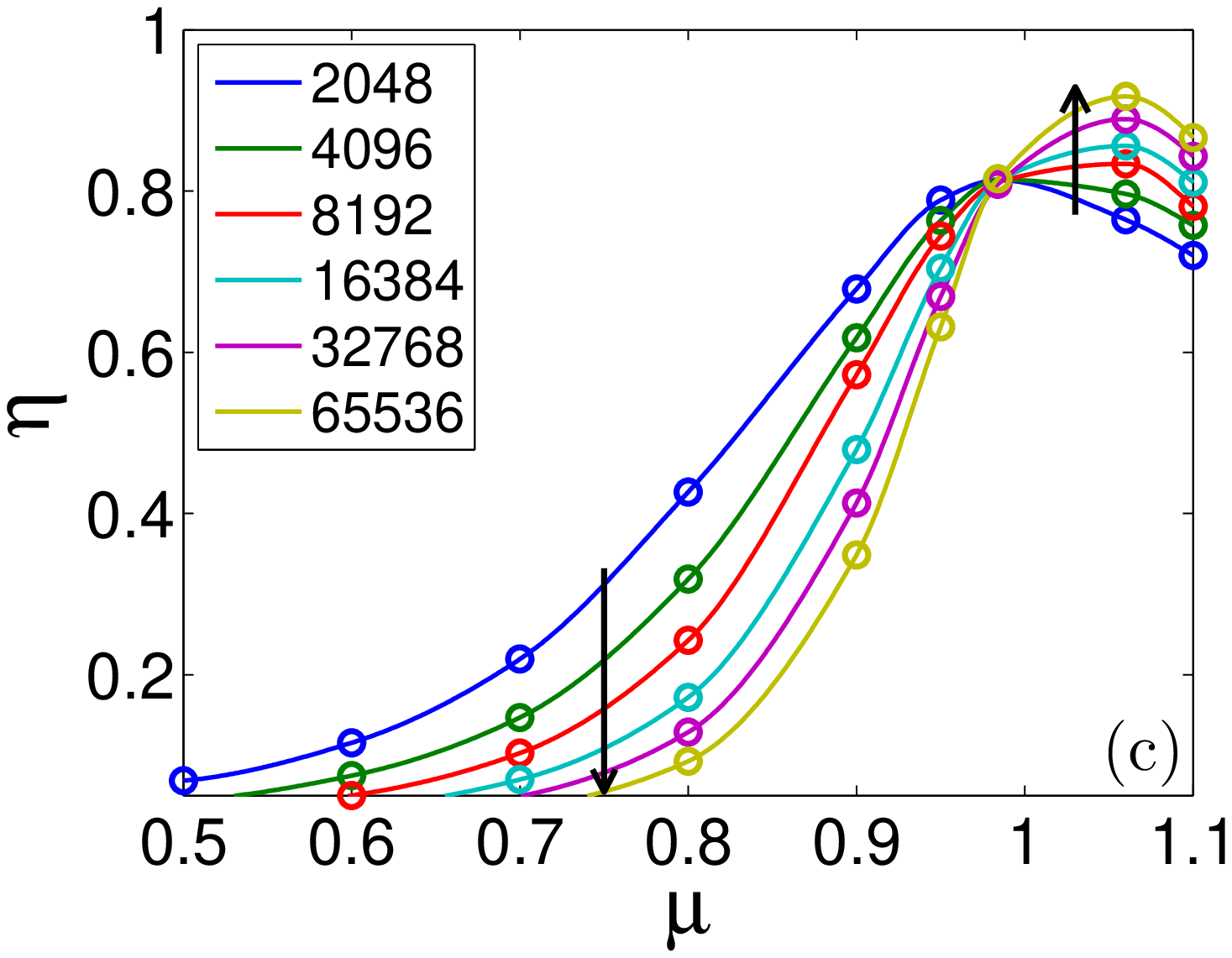}}   
  \subfigure{\includegraphics[width=0.31\textwidth]{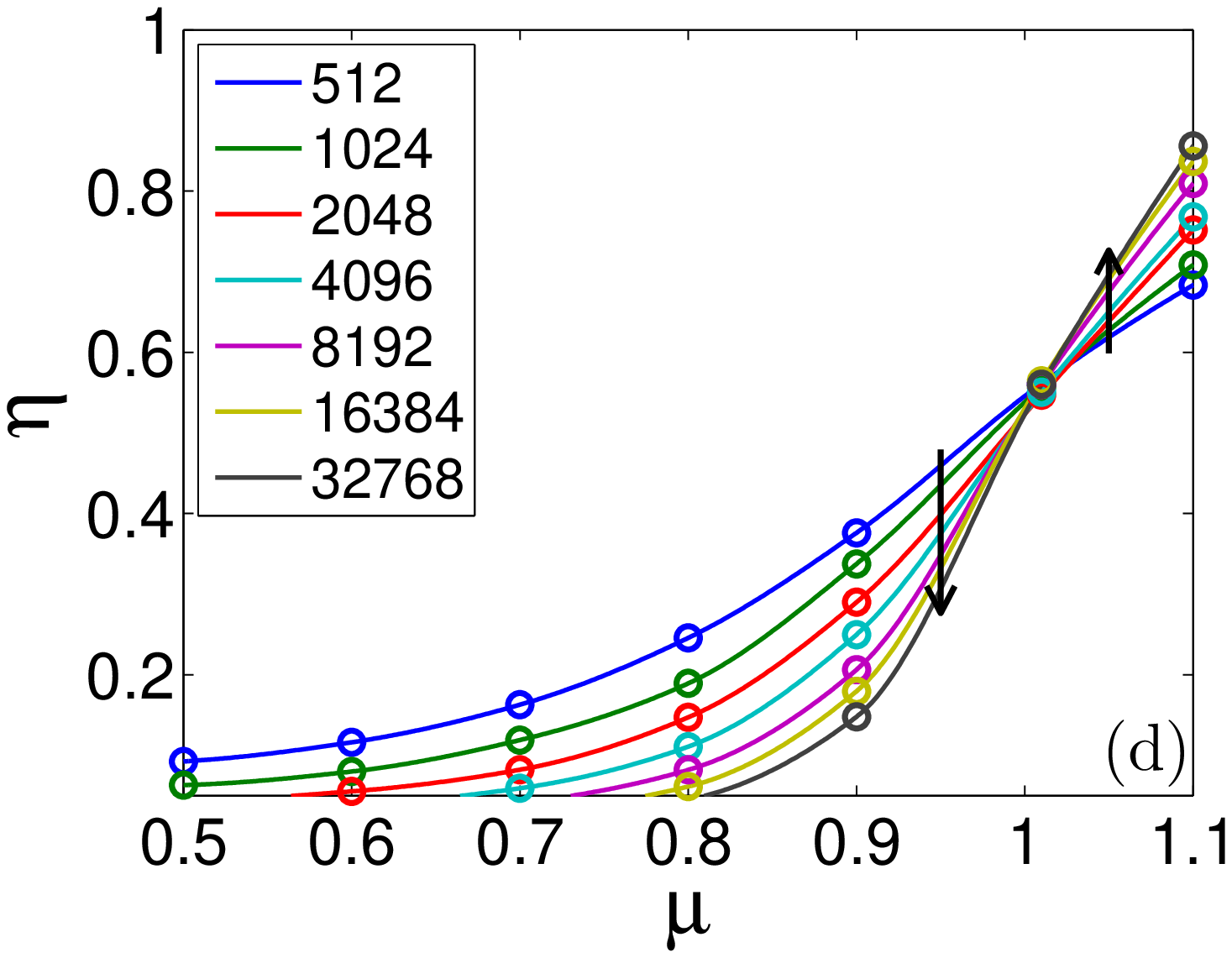}} \hspace{0.3cm} 
  \subfigure{\includegraphics[width=0.31\textwidth]{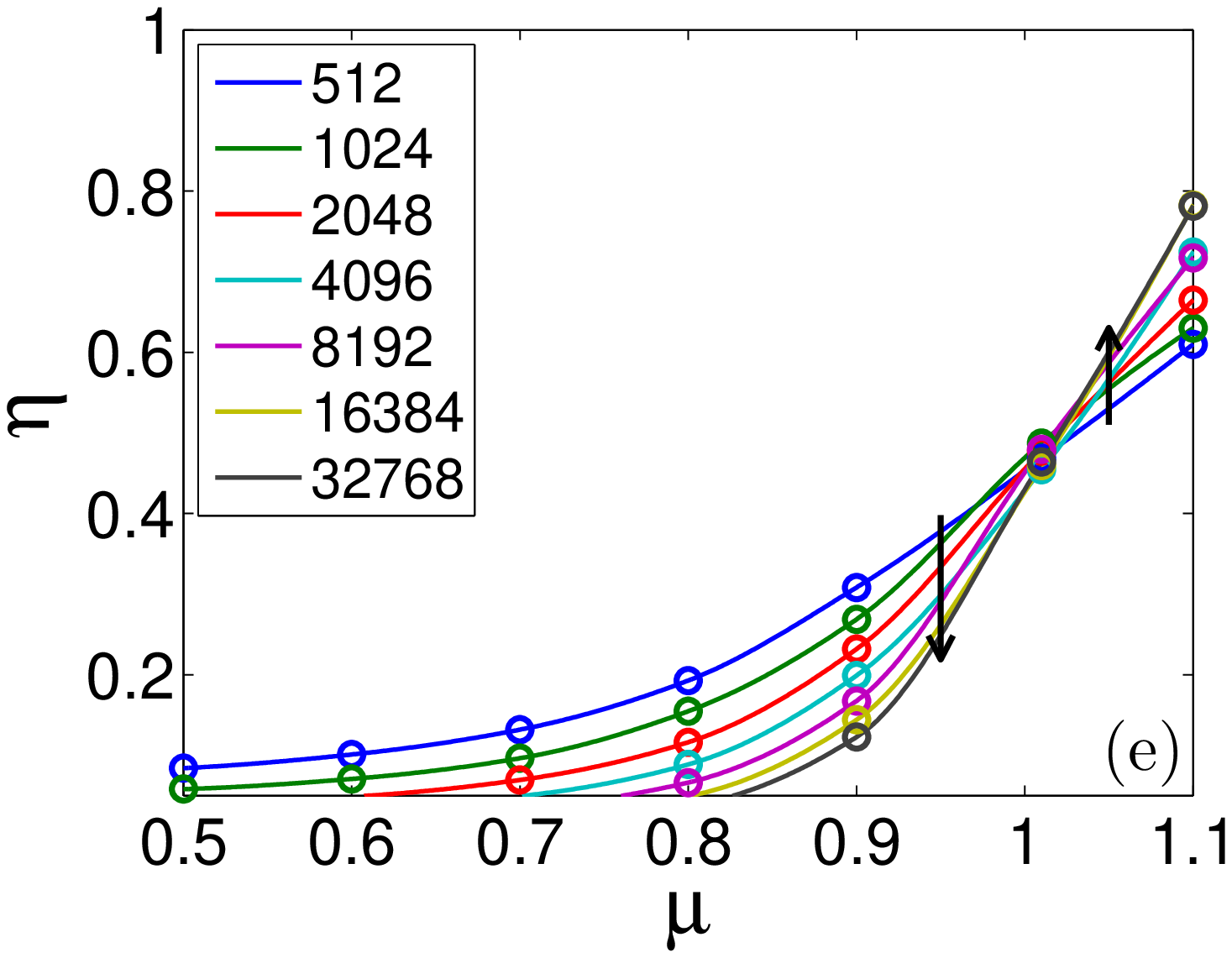}} \hspace{0.3cm} 
  \subfigure{\includegraphics[width=0.31\textwidth]{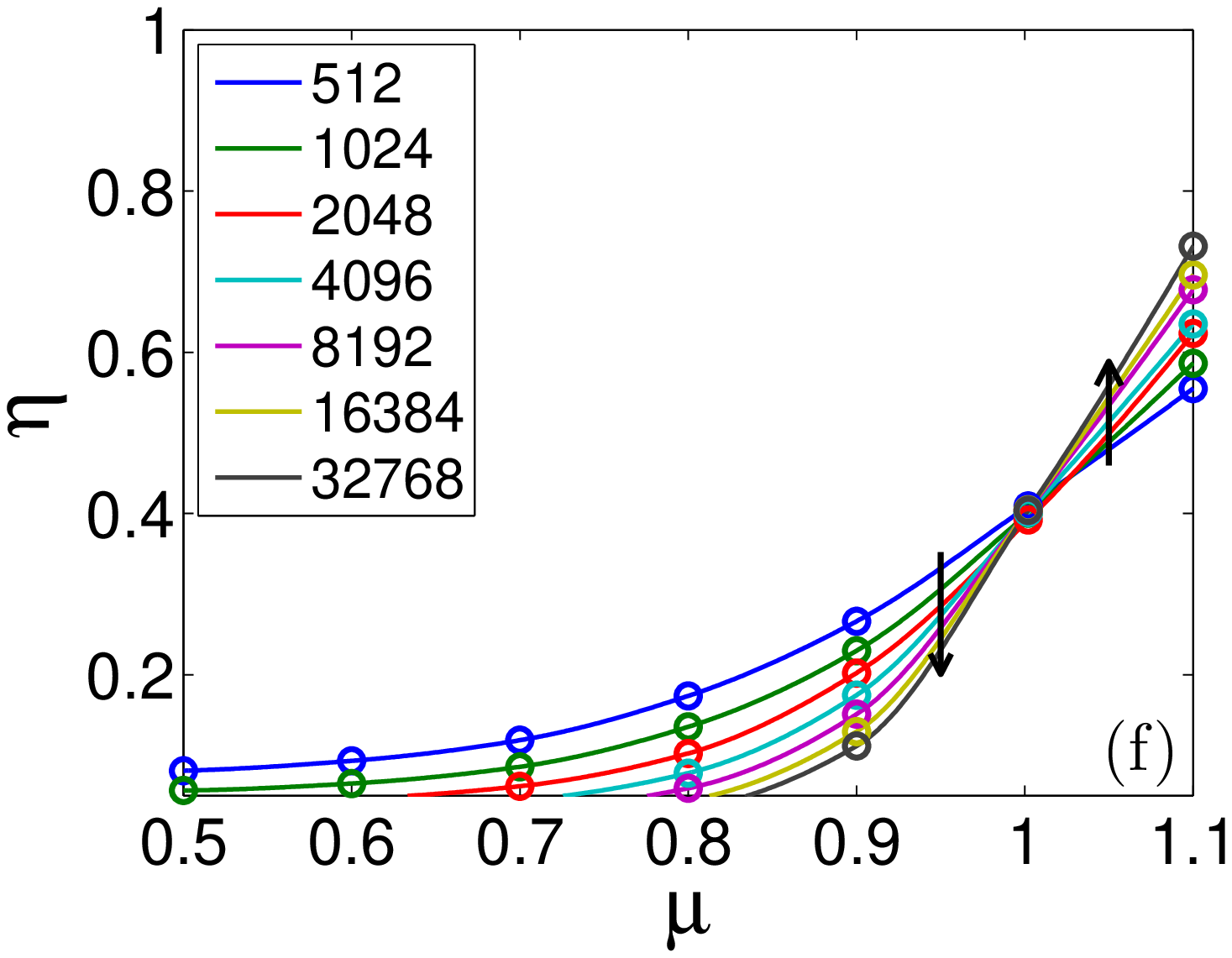}}   
\caption{(Color online) Relative fluctuation of the participation number, $\eta$, as a 
function of $\mu$ for the dPBRM model for selected values of the sparsity: (a) $\alpha=0.06$, (b)  $\alpha=0.08$, 
(c) $\alpha=0.1$, (d) $\alpha=0.4$, (e) $\alpha=0.6$, and (f) $\alpha=0.8$. 
In each panel we show curves for different network sizes (arrows indicate increasing $N$). Each point was computed from $2^{n-3}$ eigenfunctions with eigenvalues around the 
band centre with $2^{18-n}$ realisations of the dPBRM model.}
\label{Fig1}
\end{figure*}
\begin{figure*}[!htbp]
    \centering
    \subfigure{\includegraphics[width= 0.495\textwidth]{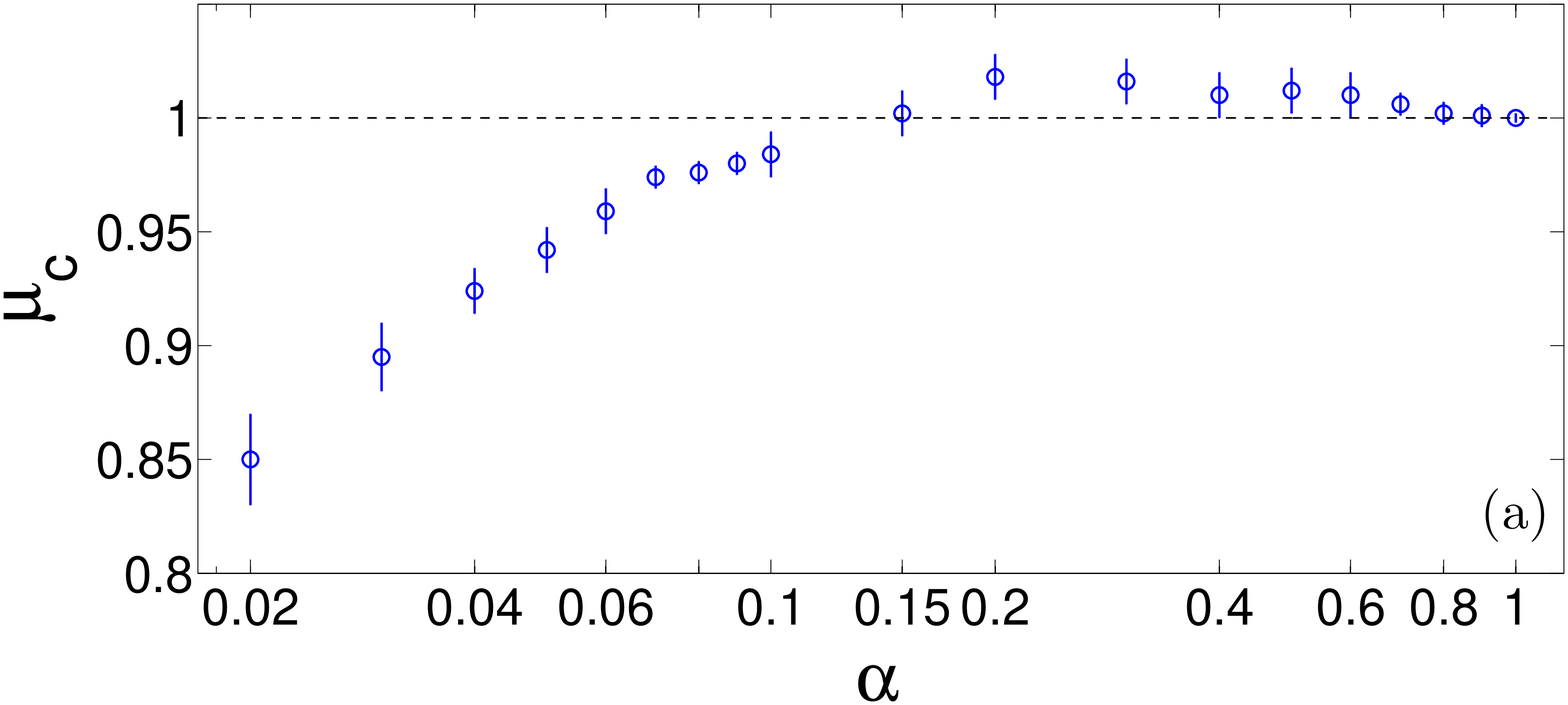}}   
    \subfigure{\includegraphics[width= 0.495\textwidth]{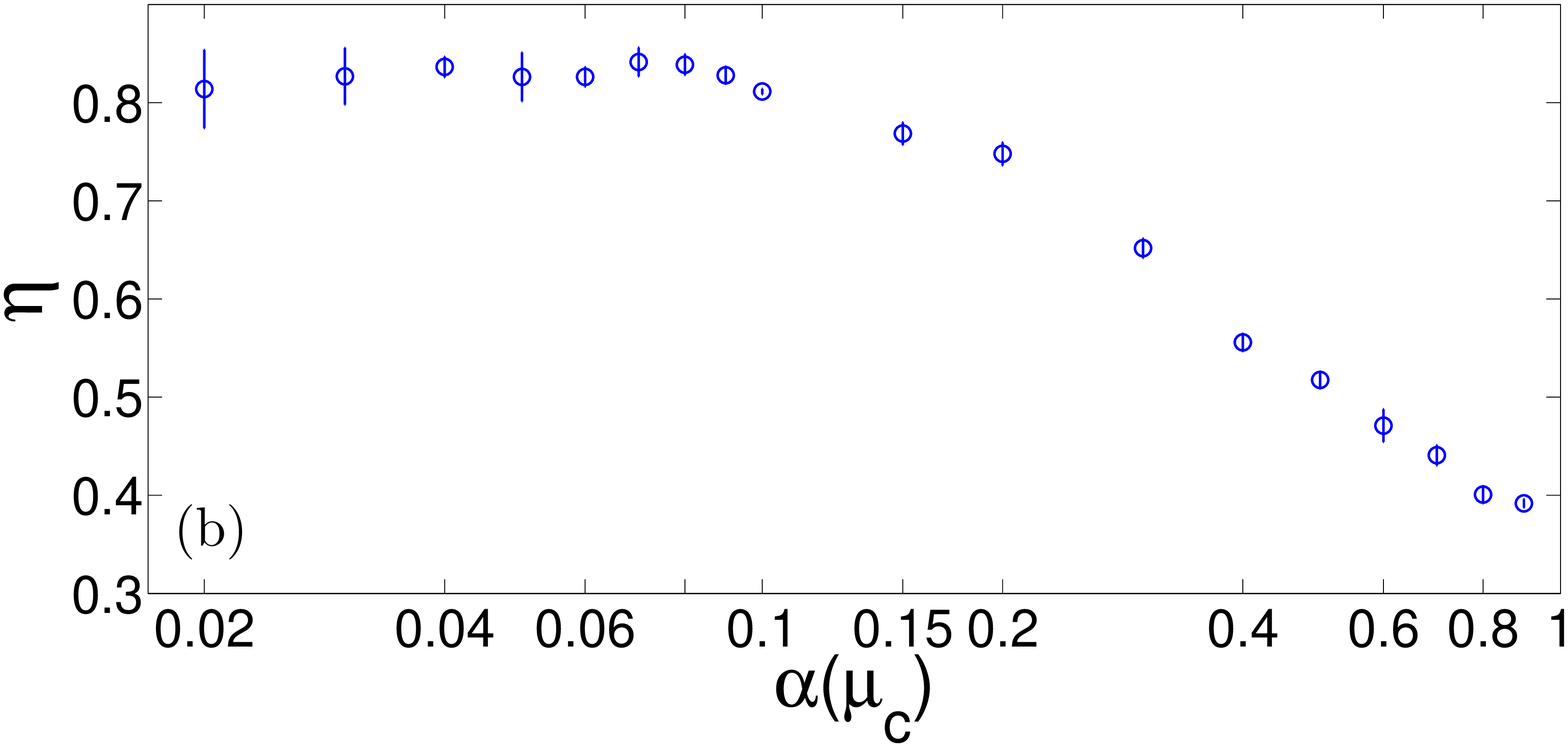}}  
\caption{(a) The critical value $\mu_c$ as a function of the sparsity $\alpha$ for the dPBRM model. Here we define error bars as the width of the $\mu$ region where the curves of Fig.~\ref{Fig1} cross. (b) Relative fluctuation of the participation number, $\eta$, at $\alpha(\mu_c)$.}
\label{Fig2}
\end{figure*}

\section{Network Eigenfunction Multifractality}
\label{sectIII}

Given an eigenfunction $\Psi$ it is a common practice (in random matrix models and 
complex Hamiltonian systems) to characterise its complexity by the use 
of the generalised participation numbers
\begin{equation}
I_q = \left( \sum_{i=1}^N |\Psi_i|^{2q} \right)^{-1} \ ,
\label{Iq}
\end{equation}
where $N$ is the corresponding matrix size. In particular, the participation number $I_2$ is roughly equal to the number of principal eigenfunction components, and therefore, is a widely accepted measure of the extension of the eigenfunction $\Psi$ in a given basis. 
Participation numbers and also inverse participation ratios (i.e.,~$(I_q)^{-1}$) have 
been already used to characterise the eigenfunctions of the adjacency matrices of random 
network models (see some examples in 
Refs.~\cite{F01,MPV07,CAS08,GGS09,JZL11,S12,PMCM13,MZN14}).

In the context of random matrix models showing the metal-to-insulator phase transition,
such as the PBRM model,
it is well established that the distribution functions of the inverse participation ratios
are scale invariant at the transition point \cite{EM00} where the eigenfunctions
are multifractal objects (see also \cite{V02}). 
The PBRM model with $\mu=1$ is at criticality, however, introducing the
sparsity $\alpha$ may relocate the metal-to-insulator transition point. Therefore,
before talking about multifractality of eigenfunctions for the dPBRM model we first have 
to be sure that the system is at criticality. Thus, to search out the critical points of 
the dPBRM model with $\alpha<1$ we use the relative fluctuation (the ratio of the standard 
deviation to the mean value) of the participation number $I_2$,
\begin{equation}
\eta = \frac{\sqrt{\left\langle (I_2)^2 \right\rangle-\left\langle I_2 \right\rangle^2}}
{\left\langle I_2 \right\rangle} \ .
\label{eta}
\end{equation}
Indeed, this quantity has been used to locate the metal-to-insulator transition point in 
random \cite{SAML08} and non-random \cite{MMD04} long--range hopping models.

In the following, we use exact numerical diagonalisation to obtain the eigenfunctions $\Psi$ of the adjacency matrices of large ensembles of random networks, represented by 
the dPBRM model (characterized by $N$ and $\alpha$).

In Fig.~\ref{Fig1} we present the curves of $\eta$ vs.~$\mu$ for the dPBRM model with sparsity $\alpha$. For a given value of the sparsity $\alpha$ we show curves corresponding to different (exponentially growing) network sizes. For all the values of $\alpha$ we consider here, $\alpha=[0.02,1]$, we observe two opposing behaviours: 
When $\mu\ll 1$ [$\mu\gg 1$] the quantity $\eta$ decreases [increases] for increasing [decreasing] network size $N$. Moreover, 
we observe that curves for different $N$ have a fixed point at $\mu=\mu_c$, revealing the invariance of $\eta$ and therefore the existence of a metal-to-insulator transition point at $\mu_c$. Then, in Fig.~\ref{Fig2}(a) we plot $\mu_c$ vs.~$\alpha$.
From this figure we can see that: for moderate sparsity, i.e., $\alpha>0.1$, $\mu_c\sim 1$; while for relatively strong sparsity, i.e, $\alpha\le 0.1$, $\mu_c$ decreases for decreasing $\alpha$.

As complementary information, in Fig.~\ref{Fig2}(b) we report the values of the relative fluctuation $\eta$ of the participation number that we found at $\mu_c$. As for $\mu_c$,
here we also observe two different behaviours for $\eta$: 
while it decreases for increasing $\alpha$ when  $\alpha>0.1$;
it is interesting to note that $\eta$ is approximately constant ($\eta\approx 0.8$) for relatively strong sparsity, $\alpha\le 0.1$.

\begin{figure*}[!htbp]
\centering	
  \subfigure{\includegraphics[width=0.31\textwidth]{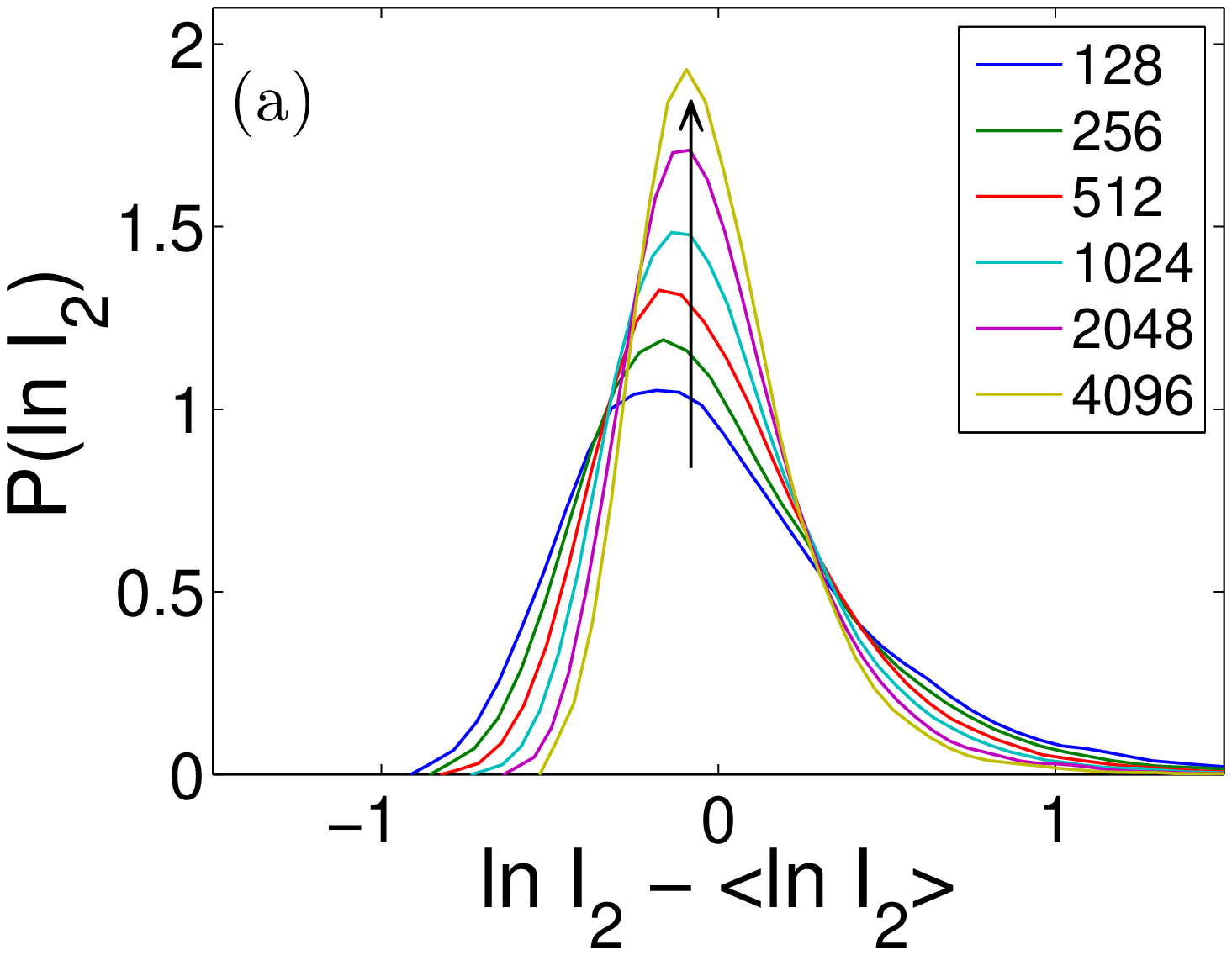}} \hspace{0.3cm}
  \subfigure{\includegraphics[width=0.31\textwidth]{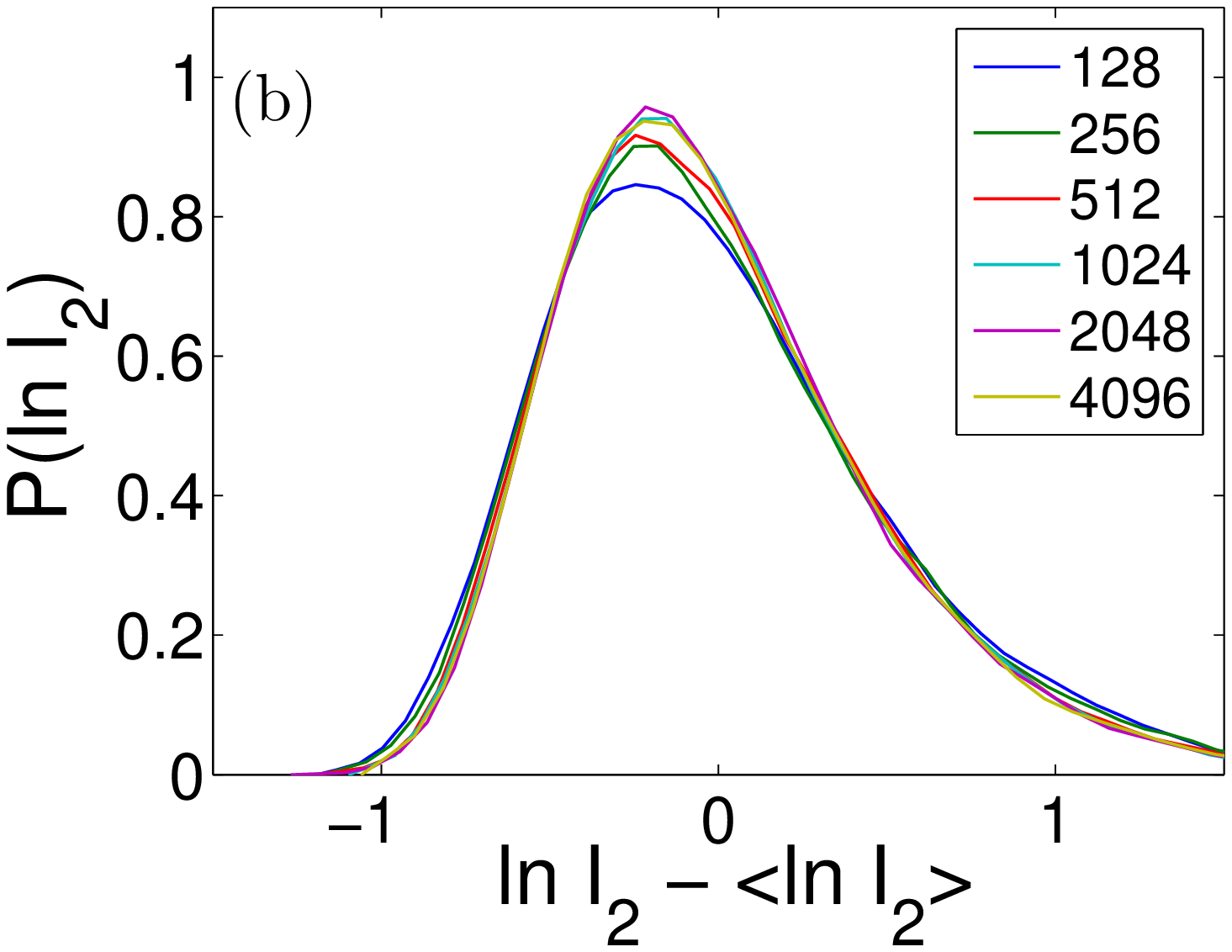}} \hspace{0.3cm} 
  \subfigure{\includegraphics[width=0.31\textwidth]{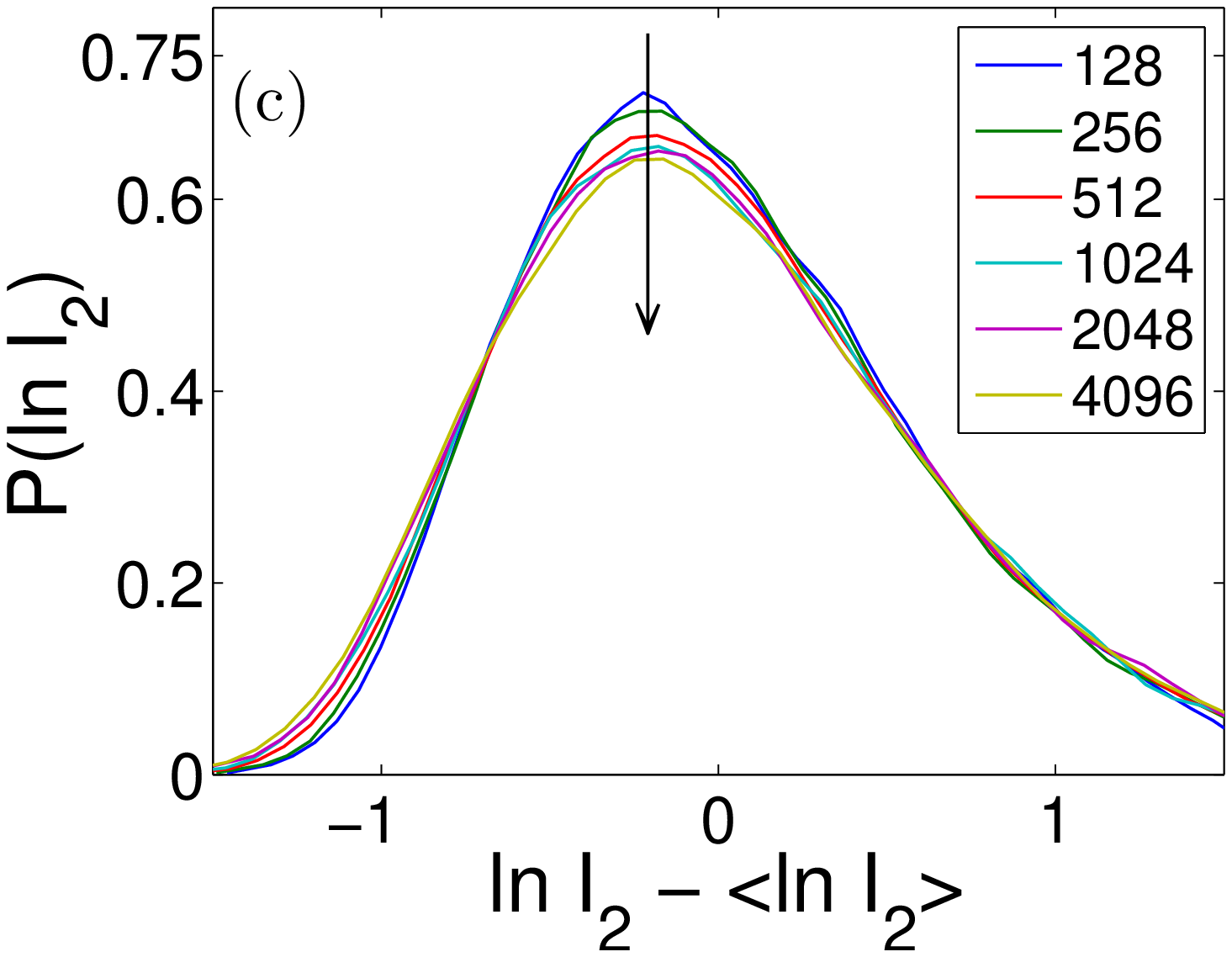}} 
  \subfigure{\includegraphics[width=0.31\textwidth]{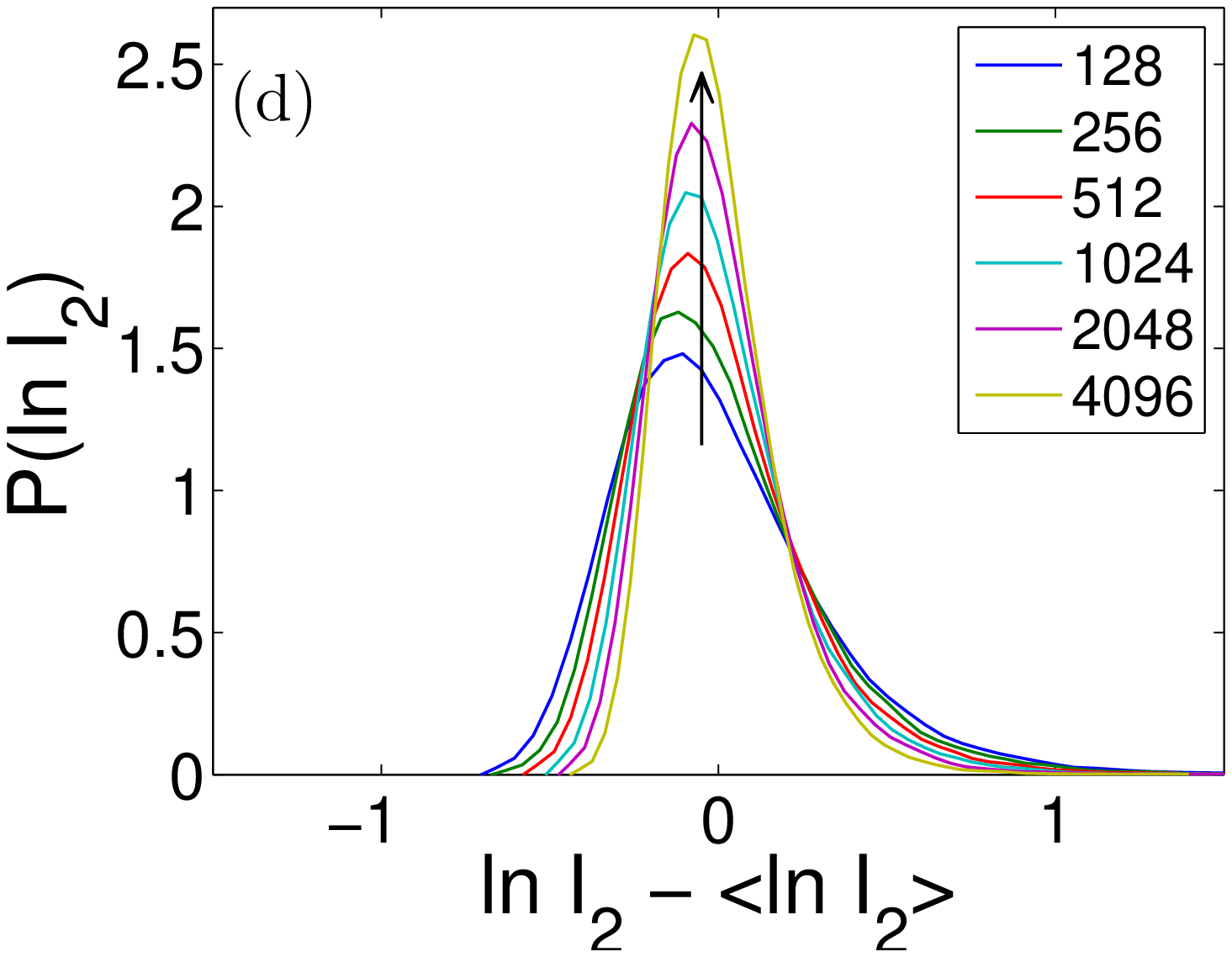}} \hspace{0.3cm}  
  \subfigure{\includegraphics[width=0.31\textwidth]{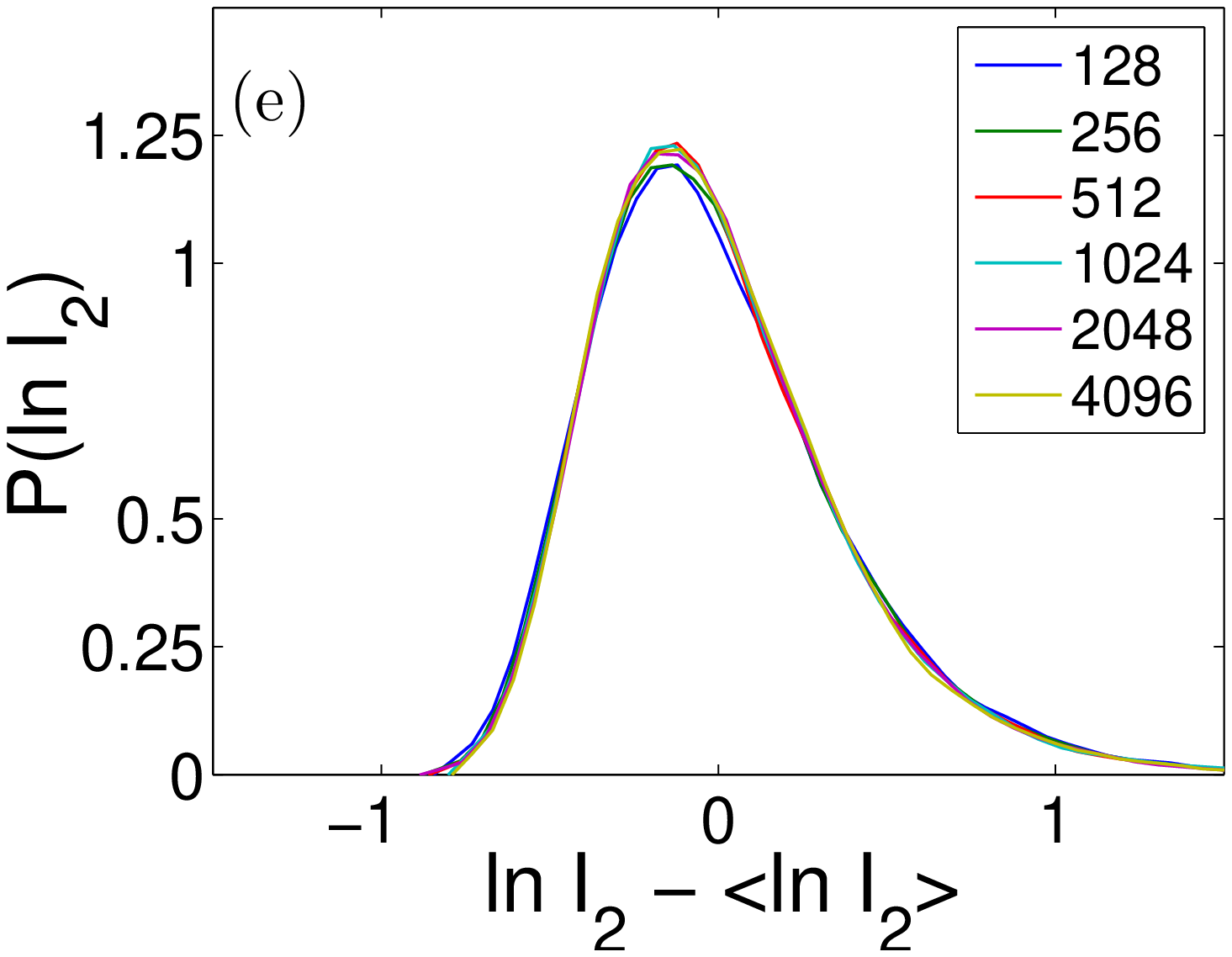}} \hspace{0.3cm} 
  \subfigure{\includegraphics[width=0.31\textwidth]{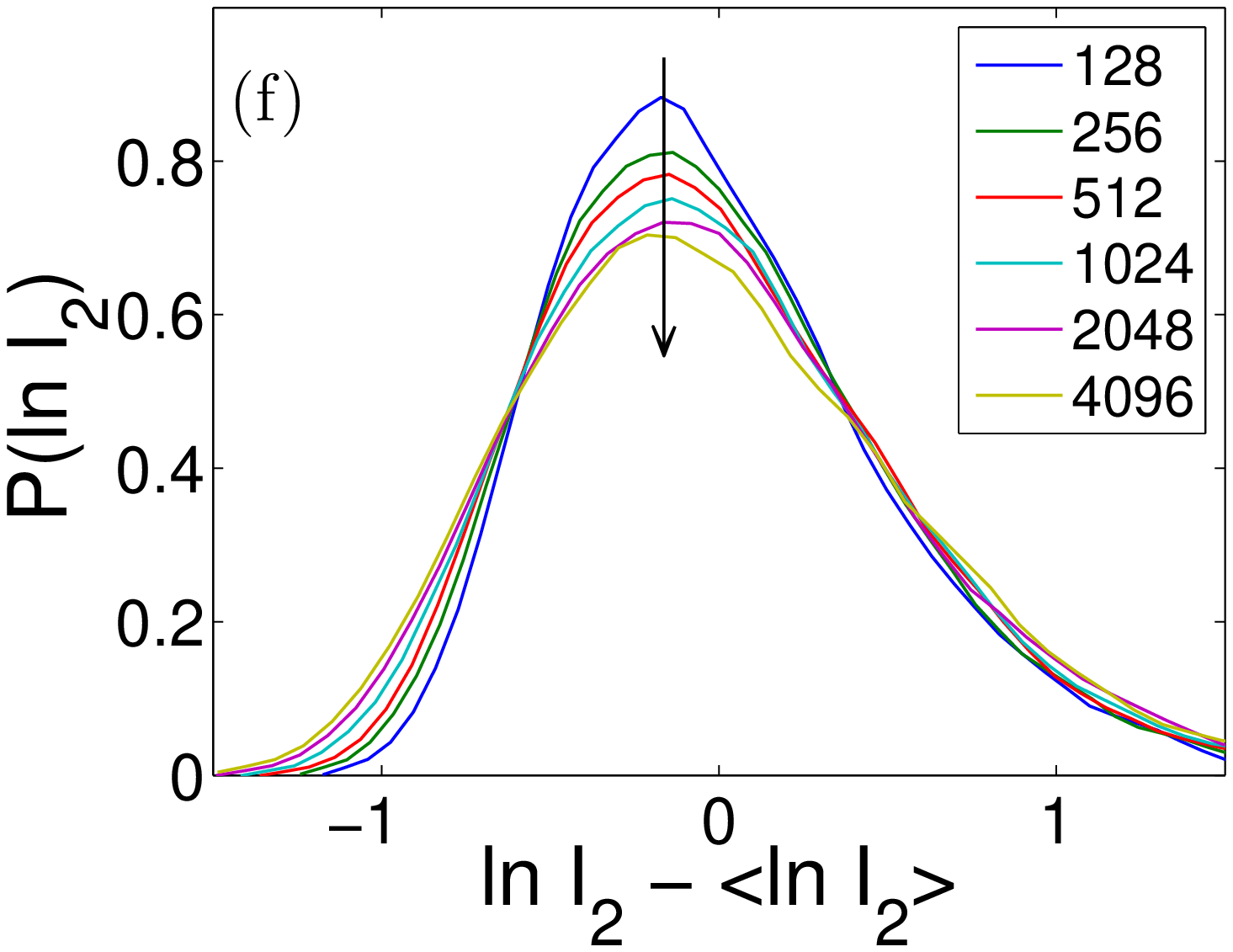}} 
  \subfigure{\includegraphics[width=0.31\textwidth]{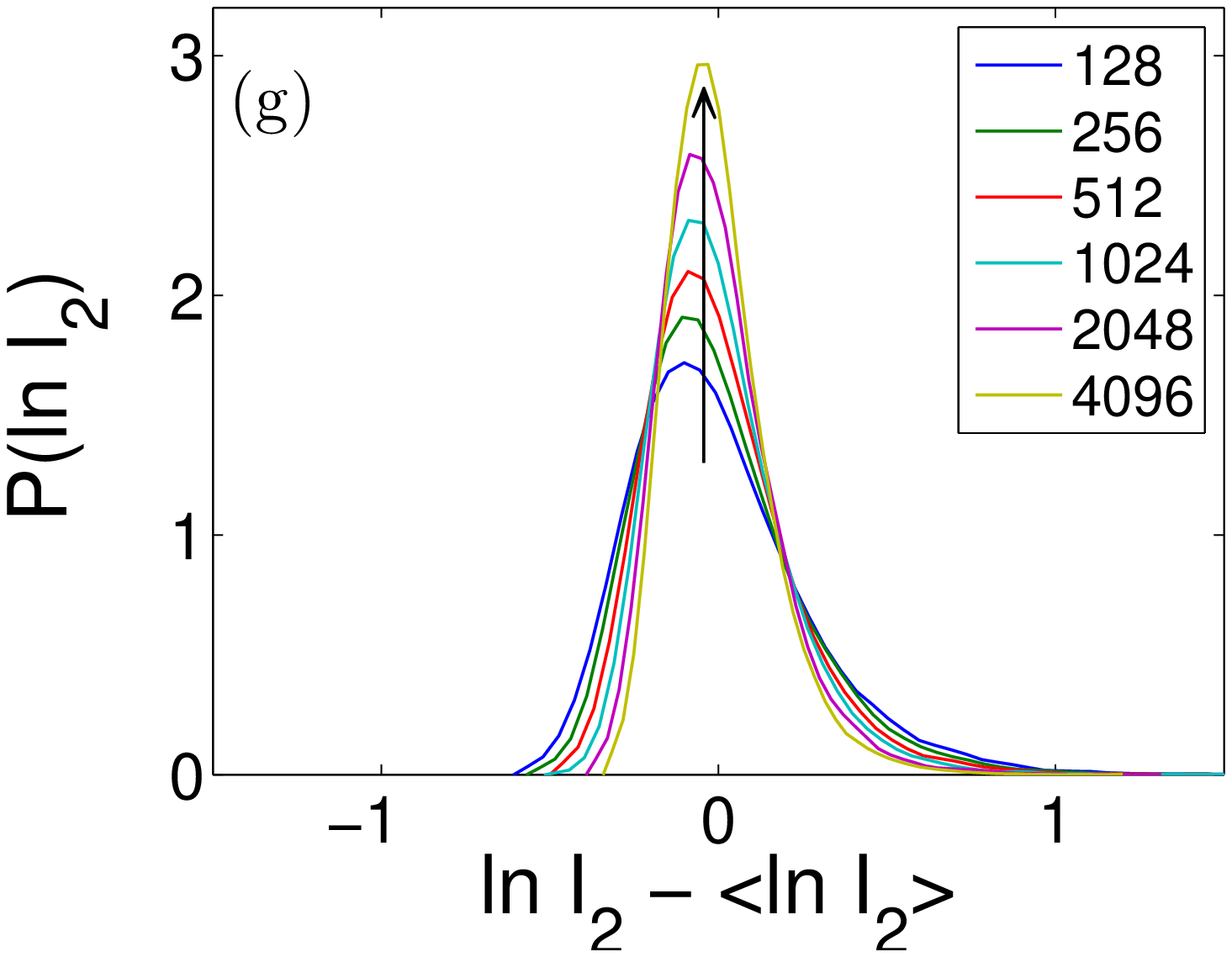}} \hspace{0.3cm}
  \subfigure{\includegraphics[width=0.31\textwidth]{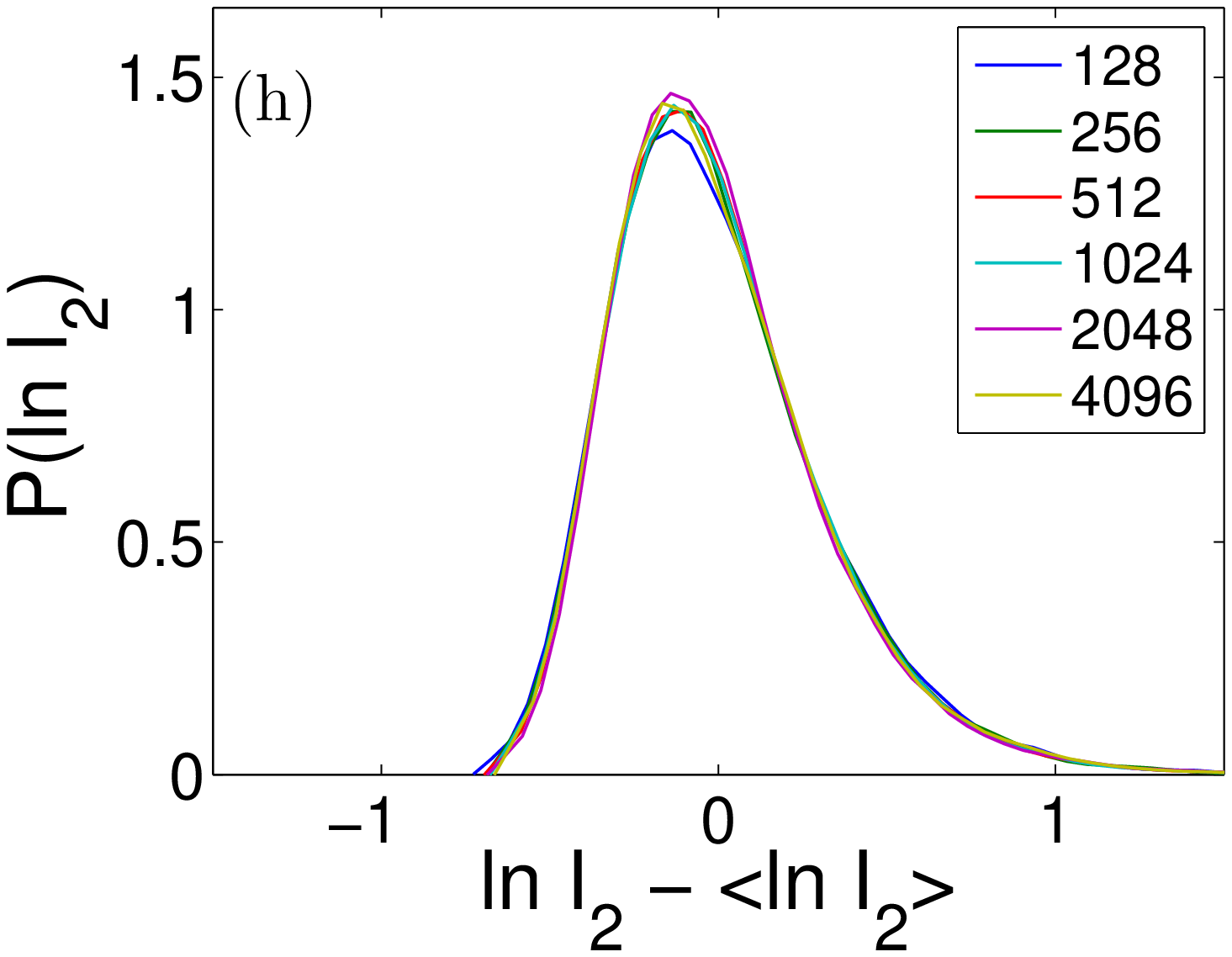}} \hspace{0.3cm}
  \subfigure{\includegraphics[width=0.31\textwidth]{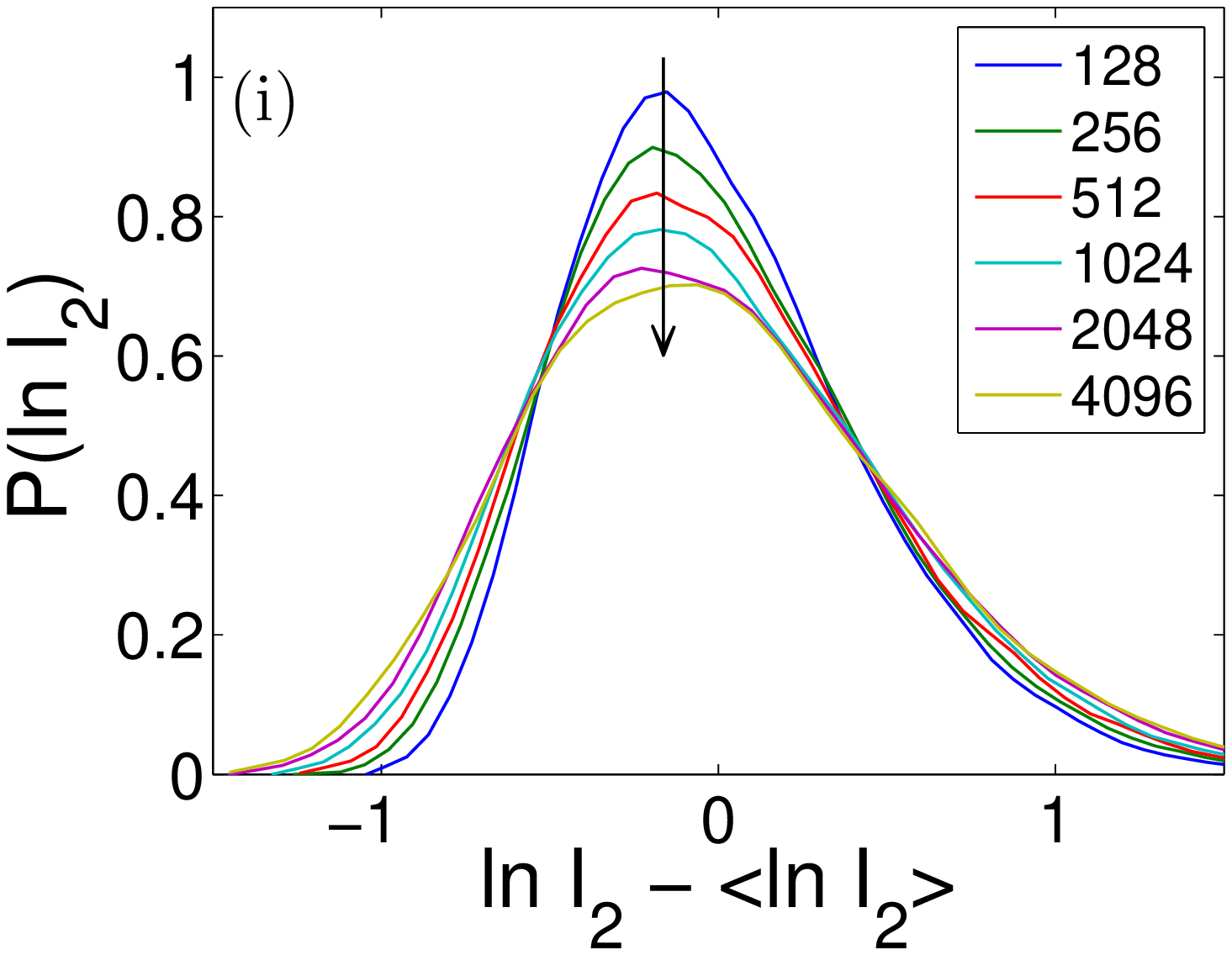}} 
\caption{(Color online) Probability distribution functions of the participation number 
$P(\ln I_2$) for the dPBRM model with sparsity $\alpha=0.3$ (upper row), $\alpha=0.6$ 
(middle row), and $\alpha=0.9$ (lower row). We used $\mu=0.9<\mu_c$ (left column), $\mu=\mu_c$ 
(middle column), and $\mu=1.2>\mu_c$ (right column). Each histogram was constructed from
$2^{15}$ data values. Arrows indicate increasing $N$.}
\label{Fig3}
\end{figure*}
\begin{figure}[!htbp]
\centering  
\includegraphics[width=0.9\columnwidth]{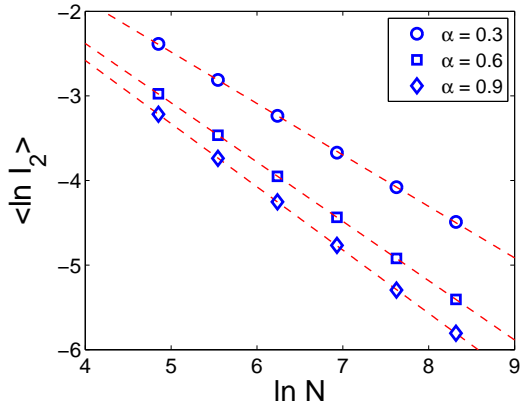}
\caption{(Color online) Logarithm of the typical participation numbers $\left\langle \ln I_2 \right\rangle$ 
as a function of the logarithm of $N$ for the dPBRM model with sparsity $\alpha$. 
Dashed lines are linear fittings to the data used to extract the following correlation dimensions: 
$D_2(\alpha=0.3)=0.6084\pm 0.0179$, $D_2(\alpha=0.6)=0.7010\pm 0.0037$, and 
$D_2(\alpha=0.9)=0.7475\pm 0.0073$.}
\label{Fig4}
\end{figure}
\begin{figure*}[!htbp]
    \centering
    \subfigure{\includegraphics[width= 0.39\textwidth]{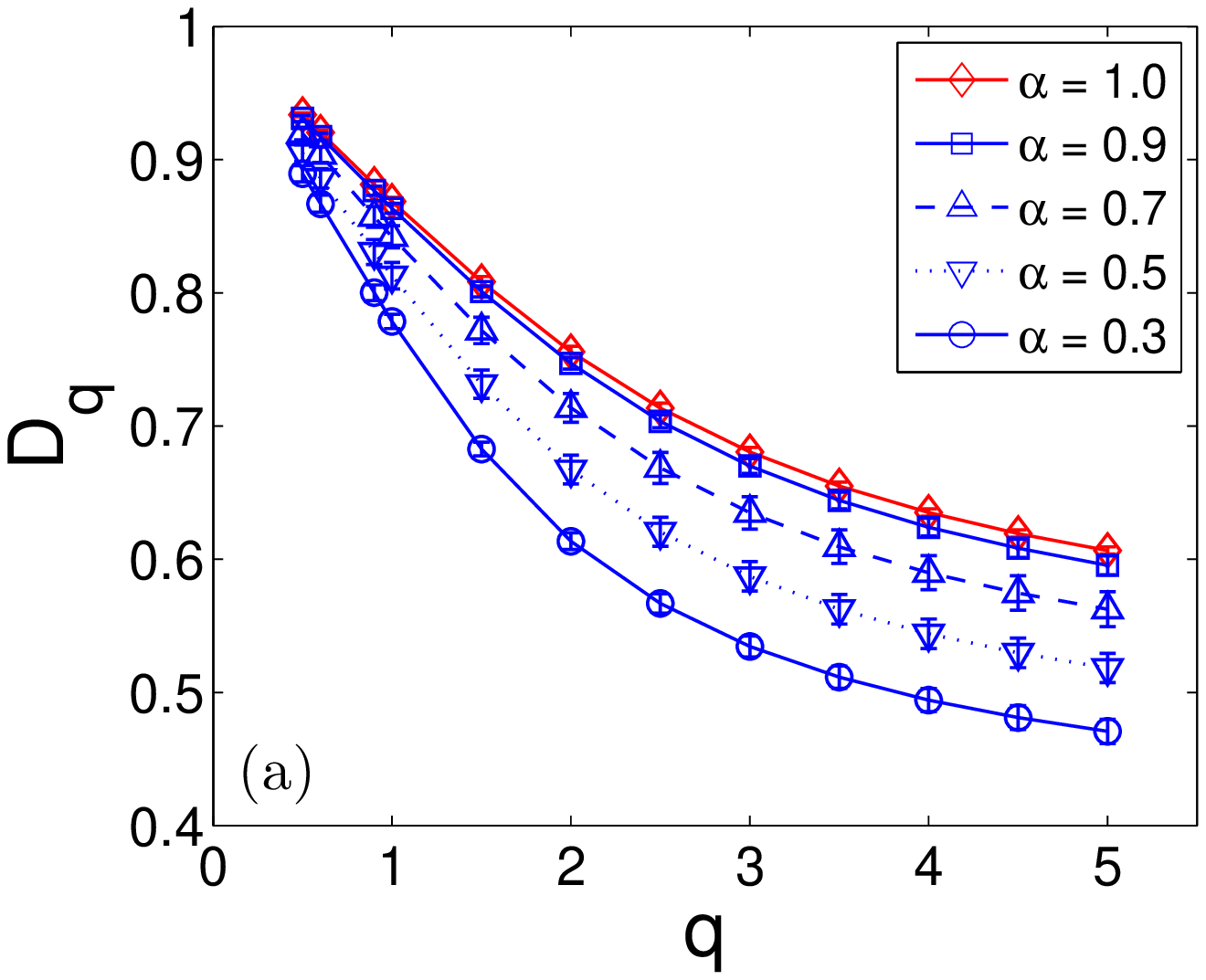}}  
    \subfigure{\includegraphics[width= 0.6\textwidth]{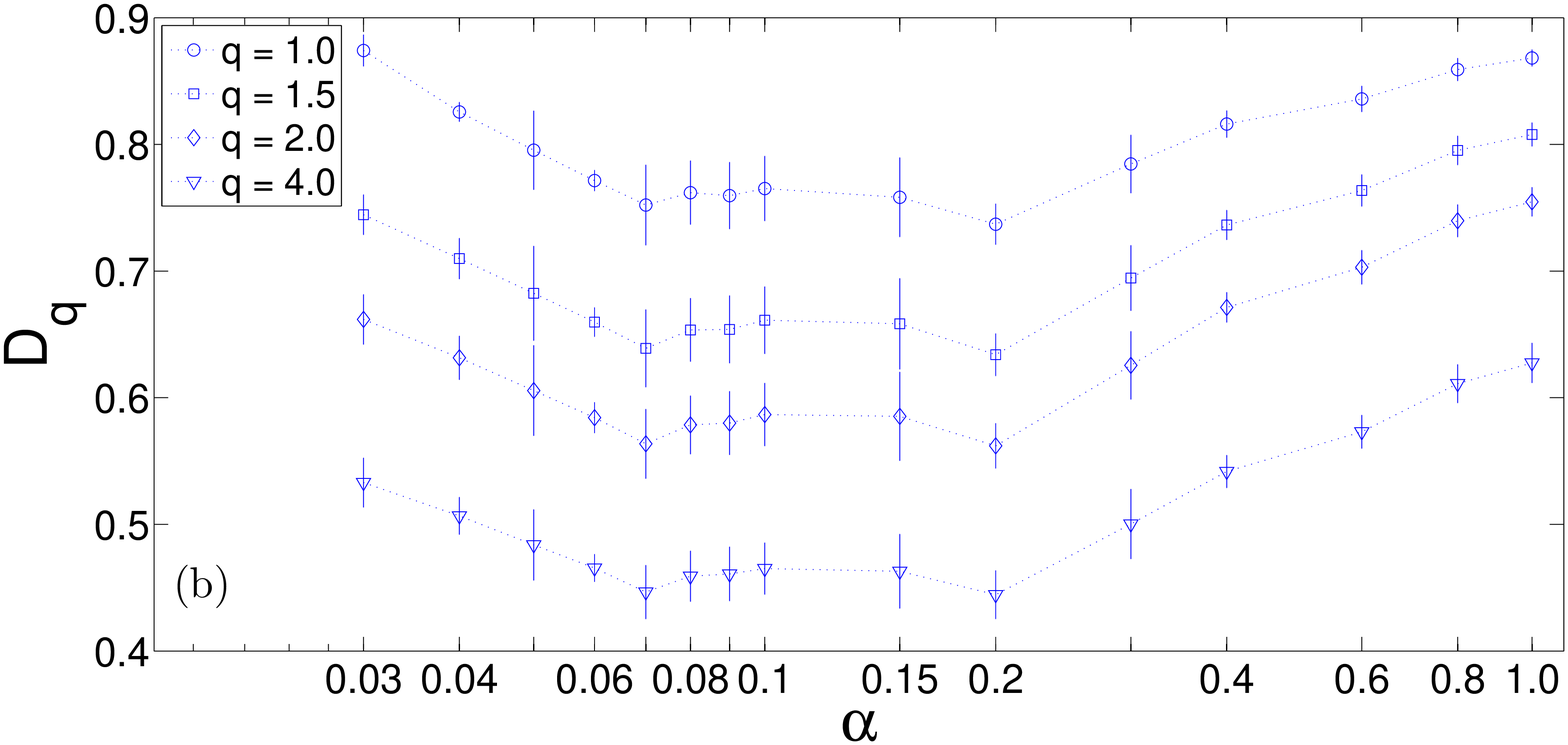}}  
\caption{(Color online) (a) Multifractal dimensions $D_q$ as a function of $q$ for the dPBRM 
model with sparsity $\alpha=0.3$, 0.5, 0.7 and 0.9 (from bottom to top). Red symbols 
correspond to the multifractal dimensions of the PBRM model (i.e., $\alpha=1$).
(b) $D_q$ as a function of $\alpha$ for selected values of $q$.}
\label{Fig5}
\end{figure*}

We stress that from Fig.~\ref{Fig1} we have located the critical points $\mu_c$ for different values of $\alpha$ (reported in Fig.~\ref{Fig2}) by the use of the invariance of the relative fluctuation of the participation number $I_2$. Moreover, we can further verify the existence of  $\mu_c$ from the invariance of the probability distribution function (PDF) of $I_2$ itself (see for example \cite{EM00,V02}). Thus, in Fig.~\ref{Fig3} we show PDFs of the participation number $P(\ln I_2)$ for the dPBRM model with sparsity $\alpha=0.3$, 0.6, and $\alpha=0.9$. For each $\alpha$ we have selected three values of $\mu$: 
$\mu=0.9<\mu_c$, $\mu=\mu_c$, and $\mu=1.2>\mu_c$. As well as for $\eta$, here we identify two behaviours for $P(\ln I_2)$ depending on whether $\mu<\mu_c$ or $\mu>\mu_c$:
When $\mu<\mu_c$ [$\mu>\mu_c$] the histograms of $P(\ln I_2)$ are narrower [wider] the larger [smaller] the network size. While, as predicted \cite{EM00,V02}, $P(\ln I_2)$ is invariant at $\mu=\mu_c$ and falls on top of a universal PDF when plotted as a function of $\ln I_2-\left\langle \ln I_2 \right\rangle$. 
We merely want to comment that for small network sizes, $N<10^3$, at $\mu=\mu_c$ we observe that $P(\ln I_2)$ evolves as a function of $N$; which is a finite size effect. 
Indeed, as clearly seen in Fig.~\ref{Fig3}, $P(\ln I_2)$ is already invariant for $N>10^3$.

Once we know the position of $\mu_c$ for the dPBRM model, we can characterise the 
multifractality of the corresponding eigenfunctions through the eigenfunction multifractal 
dimensions $D_q$, which are defined by the scaling of the typical 
participation numbers 
\begin{equation}
I_q^{\tbox{typ}} \equiv \exp \left\langle \ln I_q \right\rangle
\label{Iqtyp}
\end{equation}
as a function of $N$:
\begin{equation}
I_q^{\tbox{typ}} \propto N^{(q-1)D_q} \ .
\label{Dq}
\end{equation}
The multifractal dimensions $D_q$ can also be extracted from the scaling of 
the average participation numbers, $\left\langle I_q \right\rangle \propto N^{(q-1)D_q}$,
however here we choose to use typical participation numbers. 
We recall that for strongly localized eigenfunctions the corresponding 
$I_q^{\tbox{typ}}$ does not scale with the system size: $I_q^{\tbox{typ}}\sim 1$ and 
$D_q\to 0$ for all $q$. This situation corresponds to an insulating regime. 
While extended eigenfunctions always feel the entire system. Thus, a signature of the metallic regime is given by 
$I_q^{\tbox{typ}}\propto N$ and $D_q\to d$. Moreover, multifractal eigenfunctions 
should be described by the series of $D_q$, which are nonlinear functions of the 
index $q$.

We extract the multifractal dimensions $D_q$ from the linear fit of the logarithm of the typical participation numbers $\left\langle \ln I_q \right\rangle$ versus the logarithm of $N$ (see Eq.~(\ref{Dq})). We use $N=2^n$, $7\le n\le 12$. 
The average was performed over $2^{n-3}$ eigenfunctions with eigenvalues around the band centre with $2^{18-n}$ realisations of our dPBRM model.
As examples, in Fig.~\ref{Fig4} we present the scalings of $\left\langle \ln I_2 \right\rangle$ 
vs.~$\ln N$ for selected values of sparsity. Therefore, the correlation dimension $D_2$ is
extracted from the linear fits to the data (see dashed lines).
We note the remarkably clean linear scaling of $\left\langle \ln I_2 \right\rangle$ vs.~$\ln N$.

Finally, in Fig.~\ref{Fig5}(a) we report the multifractal dimensions $D_q$ as a function of $q$ for the dPBRM model with selected values of $\alpha$ (to avoid figure saturation). The nonlinearity of the curves $D_q$ vs.~$q$ is the signature of the multifractality of eigenfunctions of our network model. Also, as a reference, in Fig.~\ref{Fig5}(a) we include the values of $D_q$ for the PBRM model (i.e.~the values of $D_q$ for the dPBRM model with $\alpha=1$). Additionally, in Fig.~\ref{Fig5}(b) we show $D_q$ vs.~$\alpha$ for selected values of $q$, in particular we include the information dimension $D_1$ and the correlation dimension $D_2$.
From this figure we observe two behaviours: an initial decrease of $D_q$ for decreasing $\alpha$, for relatively large values of $\alpha$ ($\alpha\ge 0.2$); while, remarkably, the further decreasing of $\alpha$ (i.e., $\alpha<0.08$) makes the multifractality of the eigenfunctions of the dPBRM model to grow to values close to those for weak sparsity.

\section{Discussion and conclusions}

In this paper we consider random networks whose adjacency matrices ${\bf A}$ are represented by a sparse version of the Power--Law Banded Random Matrix (PBRM) model, therefore having a power--law structure, $A_{mn}\propto |m-n|^{-\mu}$, tuned by the parameter $\mu$ (see Eq.~(\ref{PBRMp})). We call this random network model the diluted PBRM (dPBRM) model.
We would like to emphasise that the dPBRM model belongs to the same universality class than the PBRM model, as discussed in~\cite{CRBD17} where more general long-range quantum 
hopping models in one-dimension have been studied.
 
The sparsity of the dPBRM model is driven by the average network connectivity $\alpha$: 
for $\alpha=0$ the vertices in the network are isolated and for $\alpha=1$ the network 
is fully connected. Notice that the original PBRM model is recovered for $\alpha=1$,
which is known to have multifractal eigenfunctions at the critical value $\mu=\mu_c=1$
where a metal-to-insulator phase transition takes place.
Here, we show that the dPBRM model exhibits a critical value $\mu_c\equiv\mu_c(\alpha)$ 
for $\alpha<1$, as reported in Fig.~\ref{Fig2}. 
Moreover, we found that $\mu_c\sim 1$ for $\alpha>0.1$; while 
for relatively strong sparsity, $\bra k \ket\ll N$ or $\alpha\ll 1$ (since $\bra k \ket\equiv \alpha N$,
where $\bra k \ket$ is the average degree), $\mu_c$ decreases for decreasing $\alpha$. 

In addition, we demonstrate the multifractality of the eigenfunctions of our random 
network model at $\mu_c$ by the calculation of the corresponding multifractal dimensions 
$D_q$. Indeed, we observed from Fig.~\ref{Fig5} that the multifractality of the 
eigenfunctions of the dPBRM model can be effectively tuned by the average network 
connectivity $\alpha$.

We emphasize that the calculation of $D_q$ from the finite network-size scaling of the typical eigenfunction participation numbers, see Eq.~(\ref{Dq}), is equivalent to a standard box covering algorithm (where the network size $N$ works as the box size). However, due to the normalised nature of the eigenfunctions, the scaling of $\left\langle \ln I_q \right\rangle$ vs.~$\ln N$ is very stable, as clearly shown in Fig.~\ref{Fig4}, providing quite precise values of multifractal dimensions.

Our approach may be used to investigate the multifractality of eigenfunctions in other random network models. 
Indeed, similar studies have been already performed to explore the multifractality of eigenfunctions of the Anderson 
model on Cayley trees (AMCT)~\cite{STM17,MG11,TM16} and random graphs~\cite{GGG17}.
It is relevant to stress that there are three important differences between the network model studied here and the 
AMCT studied in Refs.~\cite{STM17,MG11}:
(i) Cayley trees have a fixed degree (the AMCT in~\cite{STM17,MG11} is characterized by $k=3$), while 
due to the random-network nature of the dPBRM model the degree is defined as an average quantity here.
(ii) The dPBRM model represents networks with randomly-weighted bond strengths between vertices, while 
the AMCT in~\cite{STM17,MG11} is defined as a network with constant bond strengths.
(iii) The dPBRM model possesses an infinite line of critical points characterized by the parameter $b\in(0,\infty)$
(that we did not examine here since we fixed $b=1$ in Eq.~(\ref{PBRMp}) as a representative case), whereas
the AMCT has a single critical point for a given on-site disorder strength.
Thus, even though it may be expected that the dPBRM with $\bra k \ket\approx 3$ should show similar properties
than the AMCT in~\cite{STM17,MG11}, this must be properly verified, given the differences between both models. 
Moreover, inspired by~\cite{MG11,TM19} it should also be interesting to explore the eigenfunction statistics of the 
dPBRM model off criticality, i.e., $\mu\ne 1$ in Eq.~(\ref{PBRMp}).

The relation between the fractality of networks (in networks  specifically constructed as deterministic or disordered 
fractal objects) and the (possible) fractality of the eigenfunctions of the corresponding adjacency matrix is also another 
important subject to be explored.

We would like to add that the dPBRM model, when interpreted as a model for one-dimensional quantum
chains with long--range interactions, has characteristics proper of models currently
used in the study of excitation transport~\cite{NP18}: disorder and power--law decaying bond strengths.
Furthermore, these characteristics can presumably be implemented and tuned in state-of-the-art ion-chain 
experiments; thus the dPBRM model may find applications related to quantum 
transport with high efficiencies~\cite{NP18}.

\begin{acknowledgments}

Research carried out using the computational resources of the Center for Mathematical 
Sciences Applied to Industry (CeMEAI) funded by FAPESP (Funda\c{c}\~{a}o de 
Amparo \`{a} Pesquisa do Estado de S\~{a}o Paulo, Grant No.~2013/07375-0).
J.A.M.-B. acknowledges support from VIEP-BUAP (Grant No.~MEBJ-EXC18-G), 
Fondo Institucional PIFCA (Grant No.~BUAP-CA-169), and CONACyT (Grant No.~CB-2013/220624).
F.A.R. acknowledges CNPq (Grant No.~305940/2010-4) and FAPESP (Grant No.~13/26416-9). 
D.A.VO. acknowledges CNPq (Grant 140688/2013-7) and FAPESP (Grant No.~2016/23698-1).

\end{acknowledgments}

\bibliographystyle{plainnat}

\end{document}